\documentclass[twocolumn, showkeys, amsmath, amssymb]{revtex4}

\usepackage{graphicx}
\usepackage{bm}

\begin{document}

\preprint{}

\title{
 Ordered Phase of the Dipolar Spin Ice under $[110]$-Magnetic Fields
}

\author{Shun-ichi Yoshida}
 \email{shun1@statphys.sci.hokudai.ac.jp}
\author{Koji Nemoto}
\author{Koh Wada}
 \affiliation{
 Division of Physics, Graduate School of Science,
 Hokkaido University, Sapporo 060-0810, JAPAN}

\date{\today}

\begin{abstract}
We find that the true ground state of the dipolar spin ice system under
 $[110]$-magnetic fields is the ``$Q=X$'' structure, which is consistent
 with both experiments and Monte Carlo simulations.
We then perform a Monte Carlo simulation to confirm that there exists a
 first order phase transition under the $[110]$-field.
In particular this result indicates the existence of the first order
 phase transition to the ``$Q=X$'' phase in the field above 0.35 T for
 Dy$_2$Ti$_2$O$_7$.
We also show the magnetic field-temperature phase diagram to summarize
 the ordered states of this system.
\end{abstract}


\maketitle

\begin{figure}[tp]
 \includegraphics{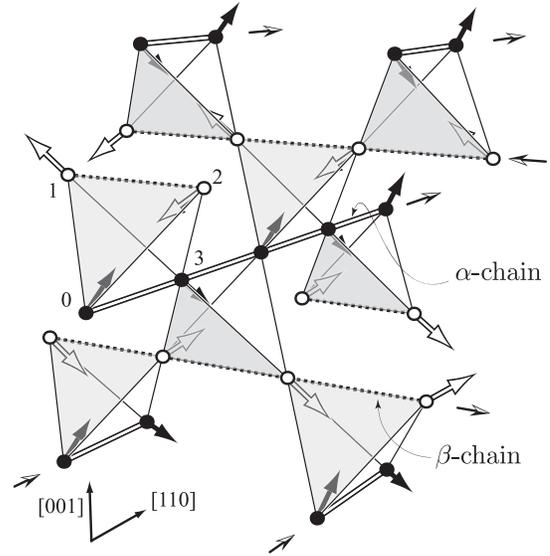}
 \caption{
 A schematic plot of the pyrochlore lattice and $Q=X$ configuration.
 The magnetic ions are located at each vertex of the tetrahedra.
 The numbers in the figure indicate the Ising axis indices $\nu=0,\,1,\,2,\,3$.
 Under the $[110]$-fields, the system is divided into two sets of spin
 chains, $\alpha$-chain (double line) and $\beta$-chain (dotted line) .
 The $[110]$-field affects spins only on the $\alpha$-chains.
 }
 \label{fig:pyrochlore}
\end{figure}

Spin ice (SI) materials such as Dy$_2$Ti$_2$O$_7$ and Ho$_2$Ti$_2$O$_7$
 have been attracting much interest as one of the ideal model materials
 of geometrically frustrated systems~\cite{Harris1997,
 Bramwell2001Science}.
In the SI materials, the magnetic moments occupy a three dimensional
 network of corner sharing tetrahedra with cubic symmetry called
 the pyrochlore lattice (FIG.\ref{fig:pyrochlore}).
Each magnetic moment is forced to point toward either of two tetrahedron
 centers by the strong crystal field along the local $\langle 111\rangle$
 axes~\cite{Siddharthan1999, Bramwell2000}.
Then ferromagnetic nearest neighbor interactions lead to a
 geometrical frustration among four spins in a tetrahedron and 
 the local ground states in each tetrahedron are six-fold degenerated.
These local ground states are determined by the rule that two spins
 point inward and two spins outward.
This ``2-in \& 2-out'' rule is called the ice rule from the analogy to
 the ice $\text{I}_\text{h}$~\cite{Harris1997}.
Under the ice rule, a number of low-lying states exist in the system and
 there remains a residual entropy almost same as the Pauling
 value~\cite{Ramirez1999,Pauling1935,Yoshida2002}.

It has been suggested 
 that the large dipole interaction is responsible for the spin ice
 behavior in the SI materials~\cite{Hertog2000}.
In the SI model with long range dipole interaction (the dipolar spin
 ice, DSI), it would be expected that the degeneracy of the ground state
 is removed by the long range nature of the dipole interaction and a long
 range order exists at low temperatures.
Hertog \textit{et al}.\ performed single spin flip Monte Carlo (MC)
 simulations of the DSI at zero field by using Ewald
 method and concluded that there is no long range order and there remains
 certain residual entropy close to the Pauling value~\cite{Hertog2000}.
Melko \textit{et al}., however, found the phase transition to the long
 range ordered $\bm{q}_\text{ord}=(0,\,0,\,2\pi/a)$ phase, where $a$ is
 the size of conventional cubic unit cell, by MC simulations employing
 the loop algorithm which improves the dynamics at low temperature.
This $\bm{q}_\text{ord}$ order was identified as the true ordered ground
 state at zero field by the Fourier transformation of the
 interaction~\cite{Gingras2001}.

Recently, the SI materials under magnetic fields along the
 $[110]$-direction have been studied by means of single crystal
 experiments~\cite{Harris1997, Fennell2002, Hiroi2003-2}.
Fennel \textit{et al}.\ performed neutron diffraction experiments of
 Dy$_2$Ti$_2$O$_7$ under the $[110]$-magnetic fields with fixed low
 temperature $\sim 70\text{ mK}$, which shows magnetic Bragg peaks at the
 $Q=0$ points and diffuse scattering at the
 $Q=X$ points~\cite{Fennell2002}.
From their scattering patterns, they found the coexistence of long-range
 ferromagnetic order on the field coupled spin chains
 (called $\alpha$-chains by Hiroi \textit{et al.}~\cite{Hiroi2003-2},
 FIG.\ref{fig:pyrochlore}) and short-range antiferromagnetic order on
 the field independent spin chains ($\beta$-chains).
They suggested that the true ground state is the $Q=X$
 structure as shown in FIG.\ref{fig:pyrochlore}~\cite{Fennell2002,
 Harris1997}, although they could not find a clear evidence for such
 ordered state.
A similar result was found in neutron diffraction experiments of
 Ho$_2$Ti$_2$O$_7$ by Harris \textit{et al}~\cite{Harris1997}.
On the other hand, Hiroi \textit{et al}.\ performed specific heat
 measurements of Dy$_2$Ti$_2$O$_7$ in the $[110]$-magnetic fields.
They found a relatively sharp peak at 1.1 K (below the broad
 peak associated with SI freezing at low fields) in the field $H \gtrsim
 0.4\text{ T}$.
They suggested that this peak comes from the freezing of $\beta$-chains
 without long range order.
They also argued the frustration between $\beta$-chains and threw
 doubt on the $Q=X$ structure as the true ground
 state~\cite{Hiroi2003-2}.
In the theoretical point of view, Melko \textit{et al}.\ performed MC
 simulations to observe a first order phase transition to the
 $Q=X$ structure~\cite{Melko2003}.
Thus the $Q=X$ structure is still controversial.

In the $[110]$-fields, there remains ground state degeneracy on the
 $\beta$-chains as far as only the nearest neighbor interaction is taken
 into account.
We expect that this degeneracy is removed by the dipole interaction.
In this paper, we give a proof that the $Q=X$ structure is
 the true ground state configuration under the $[110]$-fields.
We then perform single spin flip MC simulations
 to show that the relatively sharp specific heat peak originates from the
 first order phase transition to the $Q=X$ phase
 and to present the temperature-field phase diagram.

In the spin ice compound,
 each spin $\bm{s}_i$ is forced to align along its easy axis $\bm{n}_i$
 due to the strong crystal field, so that it is expressed with an Ising
 variable $\sigma_i(=\pm 1)$ as $\bm{s}_i = \sigma_i \bm{n}_i$,
 where $\bm{n}_i$ is one of the four distinct Ising axes $\bm{n}_\nu;\,
 \nu=0,\,1,\,2,\,3$ corresponding to each site on the basic tetrahedron
 (FIG.\ref{fig:pyrochlore}).
Then the Hamiltonian of the Ising DSI system with $N$ spins under an
 uniform external field $\bm{H}$ is given as 
\begin{align}
 &\mathcal{H}(\sigma_1,\,\sigma_2,\,\cdots,\,\sigma_N)\nonumber\\
 &=
 - p_\text{eff} \mu_\text{B} \sum_i (\bm{H} \cdot \bm{n}_i) \sigma_i
 + (J_\text{nn}+D_\text{nn}) \sum_{\langle ij \rangle}^{\text{nn}} \sigma_i \sigma_j
 \nonumber\\
 &
 + \frac{3D_\text{nn} {r_\text{nn}}^3}{5} \sum_{\langle ij \rangle}^{\text{long}}
 \frac{|\bm{r}_{ij}|^2(\bm{n}_i \cdot \bm{n}_j)
 -3(\bm{r}_{ij} \cdot \bm{n}_i)(\bm{r}_{ij} \cdot \bm{n}_j)}
 {|\bm{r}_{ij}|^5} \sigma_i \sigma_j \,,
\label{eqn:Hamiltonian_of_DipolarSpinIce}
\end{align}
where the second and the last terms represent the nearest neighbor
 interactions (both exchange $J_\text{nn}$ and dipole $D_\text{nn}$) and
 the long range part of the dipole interactions, respectively.
By using the size of unit cell $a$, the nearest neighbor distance
 $r_\text{nn}$ is written as $r_\text{nn}=a/\sqrt{8}$.
In this paper we use material parameters of Dy$_2$Ti$_2$O$_7$: 
 $J_\text{nn}=-1.24 \text{ K}$, $D_\text{nn}=2.35 \text{ K}$ and 
 $p_\text{eff}\mu_\text{B} = 10.6 \mu_\text{B} = 7.09 \text{ KT}^{-1}$~\cite{Hertog2000}.

First, we consider the effect of magnetic field along the
 $[110]$-direction without the long range interaction term.
In the $[110]$-fields, the pyrochlore lattice is divided into two sets
 of Ising spin chains named as $\alpha$ and $\beta$-chains, as
 illustrated in FIG.\ref{fig:pyrochlore}~\cite{Fennell2002, Hiroi2003-2}.
Spins on the $\nu=0,\,3$ sites compose $\alpha$-chains
 and ones on the $\nu=1,\,2$ sites $\beta$-chains.
The $[110]$-magnetic fields affect spins only on the
 $\alpha$-chains and tend to align them ferromagnetically.
Being free from the $[110]$-magnetic fields,
 spins on a $\beta$-chain also align ferromagnetically due
 to the ice rule when the $\alpha$-chains are ordered ferromagnetically.
It should be noted that each ferromagnetic $\beta$-chain has two
 possible states and the ground states are $2^{N_\beta}$-fold
 degenerated, where $N_\beta \sim N^{2/3}$ is the number of the
 $\beta$-chains.
At the ground state,
 the inter-chain configuration of $\alpha$-chains is ferromagnetic (F)
 while the one of $\beta$-chains is paramagnetic (P).
Here we introduce abbreviated notations for inter-chain configurations of the
 $\alpha/\beta$-chains for convenience.
The inter-chain configuration of the $\alpha$-chains is abbreviated as
 $\alpha$-AF if they are ordered antiferromagnetically (AF).
In this notation, the ground state configuration mentioned above is
 expressed as ``$\alpha$-F \& $\beta$-P''.
Thus there is no phase transition to the long range ordered phase without
 the long range interaction.

Now, we consider the effect of the long range interaction.
At zero field, the long range interaction removes the macroscopic
 degeneracy of the ground states and the $\bm{q}_\text{ord}=(0,\,0,\,2\pi/a)$
 configurations become the true ground states~\cite{Melko2000, Gingras2001}.
In terms of the inter-chain configuration of the $\alpha/\beta$-chains,
 these ground states are expressed as ``$\alpha$-AF \& $\beta$-AF''.
In order to show that the $Q=X$ structure, which is expressed as
 ``$\alpha$-F \& $\beta$-AF'' (see FIG.\ref{fig:pyrochlore}), is the true
 ground state under $[110]$-magnetic fields, we only consider F and AF
 inter-chain alignments for the $\alpha/\beta$-chains.
By means of the Ewald summation techniques, we calculated the
 interaction energy of the four configurations for $\bm{H}=\bm{0}$ and
 obtained
\begin{align}
&
 \frac{1}{N} E_0^{\alpha\text{-AF},\beta\text{-AF}}
 = -1.5060 \text{ K}\,,\,
 \frac{1}{N} E_0^{\alpha\text{-F},\beta\text{-F}}
 = -0.6816 \text{ K}\,,\,\nonumber\\
&
 \frac{1}{N} E_0^{\alpha\text{-F},\beta\text{-AF}}
 = \frac{1}{N} E_0^{\alpha\text{-AF},\beta\text{-F}}
 = -1.0938 \text{ K}\,.
 \label{eqn:energy0}
\end{align}
Note that the two energy differences
$E_0^{\alpha\text{-F},\beta\text{-AF}}-E_0^{\alpha\text{-AF},\beta\text{-AF}}$ and
$E_0^{\alpha\text{-F},\beta\text{-F}}-E_0^{\alpha\text{-F},\beta\text{-AF}}$
are both $0.4122\text{ K per spin}$.
This result implies that there is no interaction between the $\alpha$
 and $\beta$-chains as far as F and AF inter-chain alignment are
 concerned.
Indeed this statement is true
 since the internal fields acting on the $\beta$-chains from the
 $\alpha$-F/AF chains are canceled out each other due to the symmetry of
 the pyrochlore lattice structure.
Since ``$\alpha$-AF \& $\beta$-AF'' is the ground state at zero field,
 we can say that each set of ferromagnetic Ising spin chains tends to align
 anti-ferromagnetically.
In the $[110]$-field, the energy of the ``$\alpha$-F \& $\beta$-AF''
 configuration comes down while the energy of the ``$\alpha$-AF \& $\beta$-AF''
 configuration does not change.
So there is a ground state transition field $H_\text{c}$, at which the
 following equation holds:
\begin{equation}
  \frac{1}{N} E_0^{\alpha\text{-AF},\beta\text{-AF}}
  =  \frac{1}{N} E_0^{\alpha\text{-F},\beta\text{-AF}}
  -\frac{1}{\sqrt{6}}\, p_\text{eff}\mu_\text{B} H_\text{c} \,.
 \label{eqn:energy_aAF-bAF_aF-bAF}
\end{equation}
Using these values given in eq.(\ref{eqn:energy0}), we obtain the
 transition point $H_\text{c}=0.142\text{ T}$.
Thus we conclude that
 the ``$\alpha$-F \& $\beta$-AF'' ($Q=X$) configuration is the
 true ground state in the field $H > H_c$.

\begin{figure}[tbp]
\includegraphics[width=7cm]{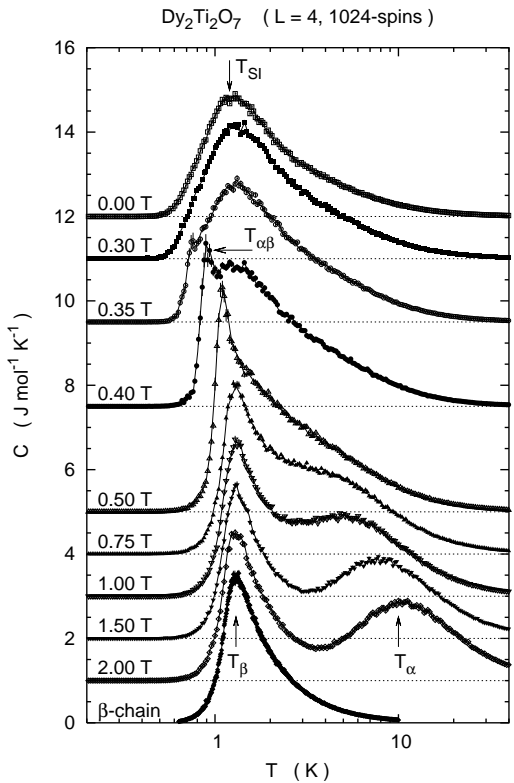}
\caption{
 Temperature dependence of the specific heat for various magnetic fields
 along the $[110]$-direction.
 Each data set has been shifted for clarity.
 The dotted line indicates the zero level for each set.
 Four characteristic peaks at
 $T_\text{SI},\,T_{\alpha\beta},\,T_\alpha,\,T_\beta$ are also
 indicated (see text).
 }
 \label{fig:T-C}
\end{figure}

Next we perform single spin flip MC simulations to show that there
 exists phase transition to the ``$\alpha$-F \& $\beta$-AF'' phase.
Our simulations were carried out on system size up to 5488 spins (cubic
 unit cell length $L=7$) with a slow cooling process.
The simulation lengths were $\sim 10^5$ MC steps per spin at each
 temperature.
In FIG. \ref{fig:T-C} we show our MC results for the temperature
 dependence of the specific heat for system size $L=4$ under
 various magnetic fields $H$.
At zero field a broad peak corresponding to the freezing
 into the SI state is found around $T_\text{SI}\simeq 1.2\text{ K}$,
 in good agreement with the previous results~\cite{Hertog2000}.
As the magnetic field increases,
 the peak at $T_\text{SI}$ grows up to $H = 0.30\text{ T}$,
 and then at $H = 0.35\text{ T}$ a small anomalous peak appears at
 $T_{\alpha\beta}$ on the low temperature side of $T_\text{SI}$, and
 this anomalous peak moves to the high temperature side.
At $H = 0.50\text{T}$, the peak at $T_{\alpha\beta}$ merges with the broad
 peak at $T_\text{SI}$ and there appears a very sharp peak,
 and then this peak is broken into two peaks at $T_\alpha$ and $T_\beta$.
The peak at $T_\alpha$ is broad and shifts to the high temperature side
 with increasing field.
On the other hand, the peak at $T_\beta$ is sharp and almost independent
 of the field.
From their field dependences, the peaks at $T_\alpha$ and at $T_\beta$
 must come from the $\alpha$ and $\beta$-chains, respectively.
To extract the contribution of the $\beta$-chains,
 we performed a MC simulation of the system only with spins on the
 $\beta$-chains.
This system equals to the DSI at the high-field limit
 since the spins on the $\beta$-chains feel no internal field from
 the $\alpha$-F chains.
The temperature dependence is also shown in the FIG. \ref{fig:T-C} and we
 see a sharp peak near $T_\beta \sim 1.3 \text{ K}$.
This sharp peak corresponds to the peak at $T_\beta$ in large but finite
 magnetic fields as expected.
These temperature and field dependences of the specific heat are also seen
 in the experiments by Hiroi \textit{et al}~\cite{Hiroi2003-2}.

\begin{figure}[tbp]
 \begin{center}
  \includegraphics[width=7cm]{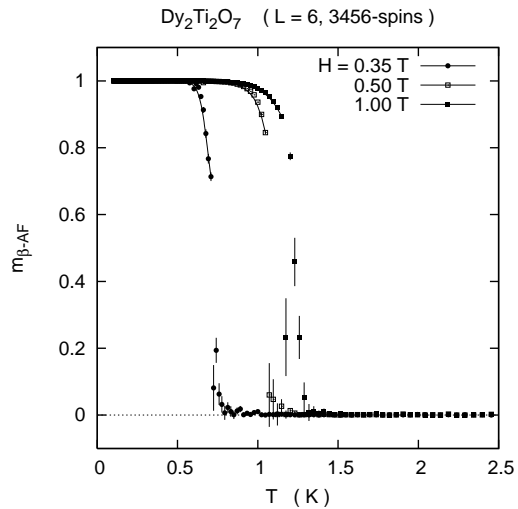}
\end{center}
 \caption{
 Temperature dependence of the $\beta$-AF order parameter for
 $[110]$-magnetic fields $H=0.35\text{ T},\,0.50\text{ T}$ and $1.00\text{ T}$\,.
 }
 \label{fig:T-betaAF}
\end{figure}

In order to consider what happens at each peak more precisely,
 we observe order parameters corresponding to the inter-chain orders.
For example,
 the order parameter for the $\beta$-AF order $m_{\beta\text{-AF}}$ is
 the $\bm{q}_\text{ord}=(0,\,0,\,2\pi/a)$ staggered magnetization of the
 $\beta$-chains.
In FIG.\ref{fig:T-betaAF} we show the temperature dependence of
 $m_{\beta\text{-AF}}$ for system size $L=6$.
\begin{figure}[tbp]
 \begin{center}
  \includegraphics[width=7cm]{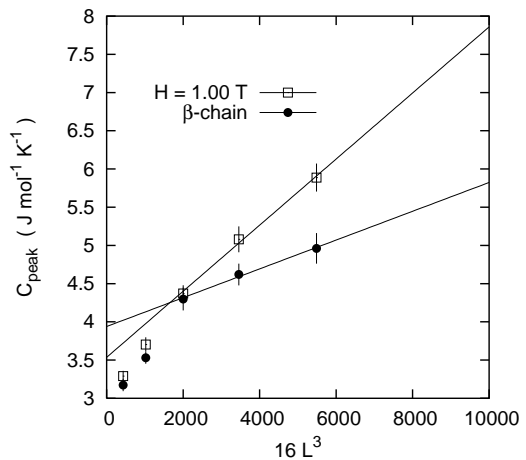}
 \end{center} 
 \caption{Finite size scaling of the peak height of the specific heat as
 a function of system size $N=16 L^3$ for L=3,\,4,\,5,\,6,\,7.}
 \label{fig:finitesizescaling}
\end{figure}
We can see the $\beta\text{-AF}$ order below $T_{\alpha\beta}$
 and $T_\beta$, and discontinuity of the order parameter
 $m_{\beta\text{-AF}}$ at these temperatures.
We also find from the observation of $m_{\alpha\text{-F}}$ that
 the field sensitive broad peak at $T_\alpha$ comes from the
 $\alpha\text{-F}$ freezing.
Therefore the low temperature anomalous peak at $T_{\alpha\beta}$ and the field
 independent peak at $T_\beta$ are due to the first order phase transition to
 the ``$\alpha\text{-F}$ \& $\beta\text{-AF}$'' ordered phase.
We also carried out a finite size scaling of the peak height $C_\text{peak}$
 for the magnetic field $H=1.00 \text{T}$ and for the $\beta$-chain
 system (FIG.\ref{fig:finitesizescaling}).
The scaling data shows that $C_\text{peak}$ is
 proportional to the system size $L^3$ for large systems ($L \ge 5$).
These results prove that the phase transition is indeed of the first order.

\begin{figure}[tbp]
 \includegraphics[width=7cm]{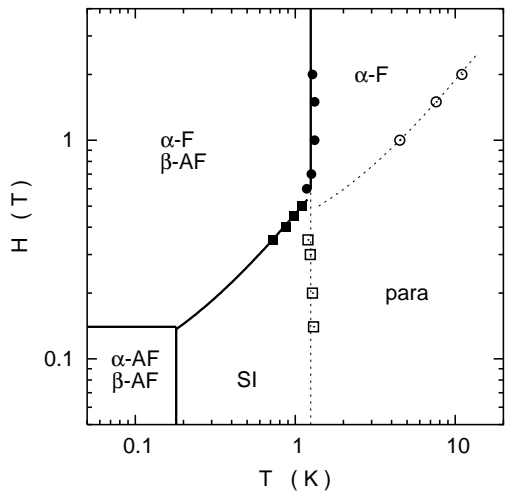}
 \caption{
 Magnetic field-temperature phase diagram for
 $\bm{H}\parallel[110]$.
 The open squares, closed squares, open circles and closed circles
 indicate the characteristic peaks $T_\text{SI}$, $T_{\alpha\beta}$,
 $T_\alpha$ and $T_\beta$, respectively.
 The solid line shows the first order phase transition and the dotted
 the crossover.
 }
 \label{fig:phase-diagram}
\end{figure}

Finally we summarize our results from the ground state analysis and the MC
 simulation as a phase diagram in FIG.\ref{fig:phase-diagram}.
The data points in the diagram indicate the specific heat peak
 temperatures $T_\text{SI},\,T_{\alpha\beta},\,T_\alpha,\,T_\beta$ in the
 MC results with single-spin-flip slow cooling process
 (FIG.\ref{fig:T-C}).
The solid line in the diagram corresponds to the first order phase
 transition.
Note that the straight lines shown in the low temperature and in the
 weak field region are expectation ones from the ground state
 transition field $H_\text{c}=0.142 \text{ T}$ and the zero field MC
 results~\cite{Melko2000}.
In the cooling process at high field region,
 spins on the $\alpha$-chains are almost fixed by the $[110]$-magnetic
 field below $T_\alpha$, while the spins on the $\beta$-chains still take
 random configurations (``$\alpha$-F'' phase).
Then at $T_\beta$ the spins on the $\beta$-chains form the $\beta$-AF order
 and the phase transition to the ``$\alpha\text{-F}$ \& $\beta\text{-AF}$''
 configuration takes place due to the ice rule and the long range dipole
 interaction. 
In the weak field region $H < 0.30 \text{ T}$, the phase transition has
 not been observed in our single spin flip MC simulation.

In conclusion, we have studied the effect of the dipole interaction on
 the DSI system under the $[110]$-magnetic fields through the ground
 state analysis and the single spin flip MC study.
We have found that the ``$\alpha\text{-F}$ \& $\beta\text{-AF}$''
 ($Q=X$) configuration is the true ground state in the
 strong field region $H > H_\text{c} = 0.142 \text{ T}$ and the first order
 phase transition from the ``$\alpha$-F'' phase to the
 ``$\alpha\text{-F}$ \& $\beta\text{-AF}$'' phase occurs.
We have presented the $[110]$-field-temperature phase diagram.
Melko \textit{et al}.\ applied the loop algorithm and the multicanonical
 MC method in order to avoid the freezing at low temperature 
 in the first order phase transition~\cite{Melko2000, Melko2003}.
Such techniques will clarify the detail of phase boundary in the low
 temperature and the weak field region, and are now under progress.

We are grateful to Z. Hiroi, K. Matsuhira, M. Ogata, and T. Sakakibara
 for useful comments and discussions. 
This research was partially supported by the Ministry of Education,
 Science, Sports and Culture, Grant-in-Aid for JSPS Fellows.

\end{document}